\documentclass[12pt]{article}
\usepackage{amsmath}
\usepackage{graphicx}
\usepackage{subfigure}
\usepackage{hyperref}

\begin{document}

\author{Ernst Trojan \and \textit{Moscow Institute of Physics and Technology} \and 
\textit{PO Box 3, Moscow, 125080, Russia}}
\title{Careful calculation of thermodynamical functions of tachyon gas}
\maketitle

\begin{abstract}
We analyze several approaches to the thermodynamics of tachyon matter. The
energy spectrum of tachyons $\varepsilon _k=\sqrt{k^2-m^2}$ is defined at $%
k\geq m$ and it is not evident how to determine the tachyonic distribution
function and calculate its thermodynamical parameters. Integrations within
the range $k\in \left( m,\infty \right) $ yields no imaginary quantities and
tachyonic thermodynamical functions at zero temperature satisfy the third
law of thermodynamics. It is due to an anomalous term added to the pressure.
This approach seems to be correct, however, exact analysis shows that the
entropy may become negative at finite temperature. The only right choice is
to perform integration within the range $k\in \left( 0,\infty \right) $,
taking extended distribution function $f_\varepsilon =1$ and the energy
spectrum $\varepsilon _k=0$ when $k<m$. No imaginary quantity appears and
the entropy reveals good behavior. The anomalous pressure of tachyons
vanishes but this concept may play very important role in the thermodynamics
of other forms of exotic matter.
\end{abstract}

\section{Introduction}

The concept of tachyon fields plays significant role in the modern research,
where they often appear in the field theory, cosmology, theory of branes and
strings with various applications \cite{S2002,BBS2003,FKS02,D1,D2}.
Tachyons, are commonly known as field instabilities whose energy spectrum is 
\begin{equation}
\varepsilon _k=\sqrt{k^2-m^2}  \label{tah}
\end{equation}
where $m$ is the tachyon mass and we use the system of standard relativistic
units $c_{light}=\hbar =\mathrm{k}_B=1$. Of course, it highly desirable to
consider such substances as an ideal gas of particles with a given energy
spectrum because this model allows to calculate all thermodynamical
parameters of exotic matter.

A system of many tachyons can be studied in the frames of statistical
mechanics \cite{M84,DHR89}, and thermodynamical functions of ideal tachyon
Fermi and Bose gases are calculated \cite{KRS07,KRS07b}. The properties of
cold tachyon Fermi gas \cite{TV2011c}, its low-temperature behavior \cite
{T2011h}, tachyonic thermal excitations \cite{TV2011d} and the hot tachyon
gas \cite{T2011i} are also investigated.

Most peculiar behavior of tachyon gas concerns that fact that the system of
tachyons may exist as a stable continuous medium when it satisfies the
causality condition 
\begin{equation}
c_s\leq 1  \label{ca}
\end{equation}
The tachyon energy spectrum (\ref{tah}) is defined at $k\geq m$, so we have
taken limits of integration in the range $m\leq k<\infty $ for all
thermodynamical quantities \cite{TV2011c}. The latter fact may result in
contradiction to the third law of thermodynamics, and we check it in the
present paper. If so, an anomalous pressure term will be necessary to avoid
this trouble. However, appearance of the anomalous pressure implies that our
previous analysis of cold tachyon Fermi gas \cite{TV2011c} is incorrect.
Either the anomalous pressure term is really present, or it is necessary to
reformulate the theory and find right definitions for the energy spectrum
and distribution function of tachyon gas.

\section{Thermodynamical functions}

Consider an ideal Fermi gas of $N$ free particles enclosed in volume $V$. At
finite temperature $T$ its statistical sum is \cite{Kapusta89} 
\begin{equation}
\ln Z=\pm \frac \gamma {2\pi ^2}V\int\limits_0^\infty \ln \left( 1\pm \exp 
\frac{\mu -\varepsilon _k}T\right) k^2dk  \label{z}
\end{equation}
where the upper and lower signs correspond to Fermi and Bose statistics,
respectively.

The distribution function is 
\begin{equation}
f_\varepsilon =\frac 1{\exp \left[ (\varepsilon _k-\mu )/T\right] \pm 1}
\label{f}
\end{equation}
depends on the single-particle energy spectrum $\varepsilon _k$ and chemical
potentia $\mu $. According to standard formulas we determine the
thermodynamical potential 
\begin{equation}
\Omega =-T\ln Z  \label{om}
\end{equation}
and the Helmholtz free energy 
\begin{equation}
F=\Omega +\mu N  \label{free}
\end{equation}
together with the particle number density 
\begin{equation}
n=\frac NV=-\frac 1V\left( \frac{\partial \Omega }{\partial \mu }\right)
_{V,T}=\frac 1V\frac{\partial \left( T\ln Z\right) _{V,T}}{\partial \mu }%
=\frac \gamma {2\pi ^2}\int\limits_0^\infty f_\varepsilon \,k^2dk  \label{n}
\end{equation}
the energy density 
\begin{equation}
E=-\frac{T^2}V\frac{\partial \left( F/T\right) _{V,\mu }}{\partial T}=\frac{%
T^2}V\frac{\partial \left( \ln Z\right) _{V,\mu }}{\partial T}+\mu n=\frac
\gamma {2\pi ^2}\int\limits_0^\infty \,f_\varepsilon \,\varepsilon _kk^2dk
\label{e}
\end{equation}
the entropy 
\begin{equation}
S=V\left( \frac{\partial P}{\partial T}\right) _{V,\mu }=V\frac{\partial
\left( T\ln Z\right) _{V,\mu }}{\partial T}=\frac{EV+P-\mu N}T  \label{s}
\end{equation}
the specific heat

\begin{equation}
C_V=T\left( \frac{\partial S}{\partial T}\right) _V=\left( \frac{\partial E}{%
\partial T}\right) _V  \label{cv}
\end{equation}
and the pressure 
\begin{equation}
P=-\frac \Omega V=\frac TV\ln Z=\pm \frac \gamma {2\pi
^2}T\int\limits_0^\infty \ln \left( 1\pm \exp \frac{\mu -\varepsilon _k}%
T\right) k^2dk  \label{pp}
\end{equation}
Integrating (\ref{p}) by parts, we have 
\begin{equation}
P=\bar P+\tilde P  \label{p}
\end{equation}
where 
\begin{equation}
\bar P=\frac \gamma {6\pi ^2}\int\limits_0^\infty f_\varepsilon \frac{%
d\varepsilon _k}{dk}k^3dk  \label{pm}
\end{equation}
and 
\begin{equation}
\tilde P=\pm \frac{\gamma T}{6\pi ^2}\left. k^3\ln \left( 1\pm \exp \frac{%
\mu -\varepsilon _k}T\right) \right| _0^\infty  \label{p0}
\end{equation}
The latter anomalous pressure (\ref{p0}) is not reflected in the specific
heat (\ref{cv}) but the entropy (\ref{s}) will be changed. It is clear that
the anomalous term (\ref{p0}) vanishes for 
ordinary subluminal particles (bradyons) with the energy spectrum 
\begin{equation}
\varepsilon _k=\sqrt{k^2+m^2}  \label{bra}
\end{equation}
and their pressure (\ref{p}) is determined by the well-known formula 
\begin{equation}
P=\bar P=\frac \gamma {6\pi ^2}\int\limits_0^\infty f_\varepsilon \frac{%
d\varepsilon _k}{dk}k^3dk  \label{ppp}
\end{equation}

According to the third law Nernst heat theorem \cite{LL48}, the entropy (\ref
{s}) must vanish at zero temperature. 
\begin{equation}
S|_{T=0}=0  \label{nern}
\end{equation}
We should always bear in mind this fact when we analyze the statistical
mechanics of tachyon ideal gas.

\section{Pressure of cold tachyon Fermi gas}

\subsection{Appearance of imaginary terms}

Consider a tachyon Fermi gas at zero temperature, when its distribution
function (\ref{f}) degenerates to the Heaviside step 
\begin{equation}
f_\varepsilon =\Theta \left( \varepsilon _F-\varepsilon _k\right)  \label{f0}
\end{equation}
where 
\begin{equation}
\mu |_{T=0}=\varepsilon _F=\sqrt{k_F-m^2}  \label{fer}
\end{equation}
is the Fermi energy and $k_F$ is the Fermi momentum. Substituting $%
f_\varepsilon $ (\ref{f0}) in (\ref{n}), (\ref{e}) and (\ref{ppp}) we find
equation for the Fermi momentum 
\begin{equation}
n=\frac \gamma {6\pi ^2}k_F^3  \label{n1}
\end{equation}
and the energy density and pressure 
\begin{equation}
E=\frac \gamma {2\pi ^2}\left( \frac{\varepsilon _Fk_F^3}4-\frac{%
m^2\varepsilon _Fk_F}8-\frac{m^4}8\ln \frac{k_F+\varepsilon _F}m\right) +E_0
\label{e1}
\end{equation}
\begin{equation}
P=\frac \gamma {2\pi ^2}\left( \frac{\varepsilon _Fk_F^3}{12}+\frac{%
m^2\varepsilon _Fk_F}8+\frac{m^4}8\ln \frac{k_F+\varepsilon _F}m\right) +P_0
\label{p1}
\end{equation}
where 
\begin{equation}
E_0=-P_0=i\frac \gamma {32\pi }\allowbreak m^4  \label{ima}
\end{equation}
The imaginary term (\ref{ima}) is constant and it is not reflected in the
sound speed 
\begin{equation}
c_s^2=\frac{dP}{dE}=\frac{dP}{dk_F}/\frac{dE}{dk_F}=\frac 13\frac{k_F^2}{%
k_F^2-m^2}  \label{c}
\end{equation}
The latter satisfies the causality condition (\ref{ca}) when 
\begin{equation}
k_F\geq \sqrt{\frac 32}m  \label{k32}
\end{equation}
Moreover, substituting (\ref{fer})-(\ref{ima}) in formula (\ref{s}), we find
that, in fact, the entropy of cold tachyon gas vanishes because 
\begin{equation}
E+P-\varepsilon_F n=0  \label{s1}
\end{equation}
that agrees with the Nernst heat theorem (third law of thermodynamics) \cite
{LL48}.

\subsection{Elimination of imaginary terms by changing limits of integration}

However, we do not know what physical meaning should pertain to imaginary
part (\ref{ima}) of the energy density (\ref{e1}) and pressure (\ref{p1}).
Thus, the above analysis is looking like no more than a mathematical trick.
The tachyon energy spectrum (\ref{tah}) is defined only at $k\geq m$, and it
is reasonable to redefine limits of integration 
\begin{equation}
\int\limits_0^\infty ...dk\rightarrow \int\limits_m^\infty ...dk  \label{lim}
\end{equation}
Substituting (\ref{lim}) in (\ref{n}), (\ref{e}) and (\ref{ppp}) we find
that the energy density and the pressure are real quantities 
\begin{equation}
E=\frac \gamma {2\pi ^2}\left( \frac{\varepsilon _Fk_F^3}4-\frac{%
m^2\varepsilon _Fk_F}8-\frac{m^4}8\ln \frac{k_F+\varepsilon _F}m\right)
\label{e2}
\end{equation}
\begin{equation}
P=\frac \gamma {2\pi ^2}\left( \frac{\varepsilon _Fk_F^3}{12}+\frac{%
m^2\varepsilon _Fk_F}8+\frac{m^4}8\ln \frac{k_F+\varepsilon _F}m\right)
\label{p2}
\end{equation}
while the Fermi momentum is defined from equation 
\begin{equation}
n=\frac \gamma {6\pi ^2}\left( k_F^3-m^3\right)  \label{n2}
\end{equation}
The sound speed is determined by the same formula (\ref{c}), and the
causality (\ref{ca}) is satisfied under the same condition (\ref{k32}). The
only trouble is that according to (\ref{s}) and (\ref{e2})-(\ref{n2}), the
entropy is finite because 
\begin{equation}
TS=V(P+E-\varepsilon _Fn)=\frac {\gamma V} {6\pi ^2}\varepsilon _Fm^3\neq 0
\label{s2}
\end{equation}
that contradicts to the third law of thermodynamics (\ref{nern}). The
violation of this law in a tachyon system had been already emphasized \cite
{M84}. Is it really so sad? Or the third law is still valid?

\subsection{Inclusion of anomalous pressure}

This problem can be resolved in the following way. Operating with limits of
integration (\ref{lim}), we should not forget to check the anomalous
pressure term (\ref{p0}) which has the form 
\begin{equation}
\tilde P=\mp \frac \gamma {6\pi ^2}Tm^3\ln \left[ 1\pm \exp \left( \frac \mu
T\right) \right] \leq 0  \label{p02}
\end{equation}
It equals to zero when $k\in \left( 0,\infty \right) $ but will be finite
when $k\in \left(m,\infty \right) $. It is always non-positive for bosons
and fermions, and for a tachyon Fermi gas at zero temperature it is
estimated as 
\begin{equation}
\tilde P_0=\tilde P|_{T=0}=-\frac \gamma {6\pi ^2}\varepsilon _Fm^3
\label{p02b}
\end{equation}
Adding (\ref{p02b}) to (\ref{p2}) we find the proper pressure of the cold
tachyon Fermi gas 
\begin{equation}
P=\frac \gamma {4\pi ^2}\left( \frac{\varepsilon _Fk_F^3}6+\frac{%
m^2\varepsilon _Fk_F}4+\frac{m^4}4\ln \frac{k_F+\varepsilon _F}m\right)
-\frac \gamma {6\pi ^2}\varepsilon _Fm^3  \label{p2b}
\end{equation}
Taking the energy density (\ref{e2}), the particle number density (\ref{n2})
and the right-defined pressure (\ref{p2b}), we see that condition (\ref{s1})
takes place, and, hence, the entropy (\ref{s}) vanishes $S=0$.

The anomalous term (\ref{p02b}) added to the pressure (\ref{p2b}) results in
sufficient changes of the tachyon gas parameters. Now the pressure $P$ never
exceeds the energy density $E$, while the sound speed 
\begin{equation}
c_s^2=\frac{dP}{dE}=\frac 13\frac{k_F^2+k_Fm+m^2}{\left( k_F+m\right) k_F}
\label{c2}
\end{equation}
is always subluminal at all $k_F\geq m$. It may seem that the previous
research of tachyon Fermi gas \cite{TV2011c,TV2011d} is wrong because we
have not taken into account the evident anomalous pressure (\ref{p02b}).
However, we need to check whether this concept is working at finite
temperature. Otherwise, we need to look for the right approach to the
statistical mechanics of tachyons.

\section{Tachyon pressure at finite temperature}

Consider the tachyon anomalous pressure (\ref{p02}) at finite temperature.
It is 
\begin{equation}
\tilde P=-\frac \gamma {6\pi ^2}m^3\left[ \mu +T\exp \left( \frac{-\mu }%
T\right) \right]  \label{p05}
\end{equation}
for a Fermi gas at low temperature $T\ll \mu $, while at zero temperature
the pressure (\ref{p05}) is reduced to (\ref{p02b}).

%
The anomalous pressure term (\ref{p05}) is incorporated in the entropy (\ref
{s}), namely 
\begin{equation}
S=\frac VT\left( E-\mu n+\bar P+\tilde P\right)  \label{sb}
\end{equation}
At zero temperature we have $S|_{T=0}$ because $\mu |_{T=0}=\varepsilon _F$
and 
\begin{equation}
E|_{T=0}-\varepsilon _Fn+\bar P|_{T=0}+\tilde P|_{T=0} =0  \label{sb2}
\end{equation}
according to (\ref{e2}), (\ref{n2}) and (\ref{p2b}).

Subtracting (\ref{sb2}) from (\ref{sb}) we have 
\begin{equation}
S=\bar S+\frac VT\left( \tilde P-\tilde P_0\right)  \label{s3}
\end{equation}
where 
\begin{equation}
\bar S=\frac VT\left( E-E|_{T=0}-\mu n+\varepsilon _Fn+\bar P-\bar
P|_{T=0}\right)  \label{s4}
\end{equation}
is the entropy of tachyon gas when the anomalous pressure (\ref{p0}) is not
taken into account. It is clear that $\bar S|_{T=0}\neq 0$.

The chemical potential of tachyon Fermi at low temperature acquires a
quadratic dependence on temperature \cite{T2011h} 
\begin{equation}
\mu =\varepsilon _F\left( 1-\frac{\pi ^2}6\frac{k_F^2+\varepsilon _F^2}{%
k_F^2\varepsilon _F^2}T^2\right)  \label{mu}
\end{equation}
and the energy density and pressure are expanded in a power series of $T $: 
\begin{equation}
E=E|_{T=0}+O\left( T^2\right) \qquad \bar P=\bar P|_{T=0}+O\left( T^2\right)
\label{mu1}
\end{equation}
It results in a linear dependence of the entropy density on temperature 
\begin{equation}
\bar S=\frac{\gamma V}6\varepsilon _Fk_FT  \label{s5}
\end{equation}

Appearance of anomalous pressure $\tilde P$ results in additional
contribution to the entropy (\ref{s3}). Substituting (\ref{mu}) in (\ref{p05}%
), we find a low temperature expansion of the anomalous pressure 
\begin{equation}
\tilde P=\tilde P_0+\frac{\gamma m^3}6\left[ \frac{\varepsilon _F}6\frac{%
k_F^2+\varepsilon _F^2}{k_F^2\varepsilon _F^2}T-\frac 1{\pi ^2}\exp \left( 
\frac{-\varepsilon _F}T\right) \right]  \label{p08}
\end{equation}
Thus, substituting (\ref{s5}) and (\ref{p08}) in (\ref{s3}) we find the
proper entropy 
\begin{equation}
S=\frac{\gamma m^3V}{6\pi ^2}\left[ \pi ^2\left( \frac{k_F}{m^3}+\frac 16%
\frac{k_F^2+\varepsilon _F^2}{k_F^2\varepsilon _F^2}\right) \varepsilon
_FT-\exp \left( \frac{-\varepsilon _F}T\right) \right]  \label{sg}
\end{equation}

\begin{figure}[tbp]
\caption{Solution of inequality (\ref{in}: $T/\varepsilon _F$ vs. $\lambda$. 
}
\label{ano1}{\includegraphics[scale=0.8]{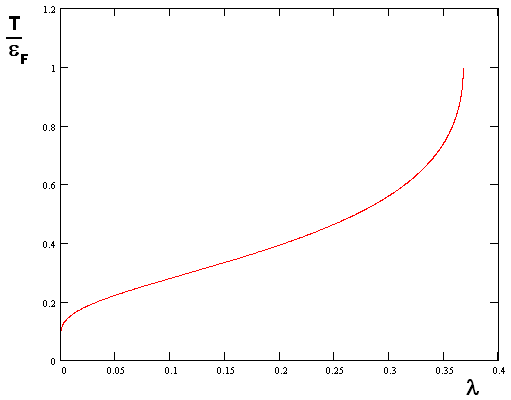}}
\end{figure}

Consider function 
\begin{equation}
g\left( T\right) =\lambda \frac T{\varepsilon _F}-\exp \left( \frac{%
-\varepsilon _F}T\right)  \label{fu}
\end{equation}
where 
\begin{equation}
\lambda =\pi ^2\left( \frac{k_F}{m^3}+\frac 16\frac{k_F^2+\varepsilon _F^2}{%
k_F^2\varepsilon _F^2}\right) \varepsilon _F^2  \label{lam}
\end{equation}
This function is negative $g\left( T\right) <0$ when 
\begin{equation}
\lambda <\frac{\varepsilon _F}T\exp \left( \frac{-\varepsilon _F}T\right)
\label{in}
\end{equation}
where $\lambda \ll 1$ because $T\ll \varepsilon _F$. The dependence of
critical ratio $T/\varepsilon _F$ vs. $\lambda $ is given in Fig.~\ref{ano1}%
. Inequality (\ref{in}) is realized when $\lambda <1/e$. Parameter $\lambda $
(\ref{lam}) can be arbitrary small when $k_F\rightarrow m$ and $\varepsilon
_F\rightarrow 0$. Hence, function $g$ (\ref{fu}) and the entropy $S$ (\ref
{sg}) will become negative at finite temperature when $k_F\rightarrow m$,
although $S=g\left( T\right) =0$ when $T\rightarrow 0$.

Now the definition of the entropy (\ref{sg}) contradicts to the laws of
thermodynamics and it implies that the concept of anomalous pressure term (%
\ref{p02}) is not working here. The problem is hidden in our definition of
the distribution function (\ref{f}) and the limits of integration (\ref{lim}%
).

\section{Discussion: right thermodynamical functions of tachyons}

When we calculate the thermodynamical functions of a cold Fermi gas with the
tachyon energy spectrum (\ref{tah}), the energy density (\ref{e1}) and
pressure (\ref{p1}) may include imaginary parts. Imaginary terms does not
appear if we change limits of integration (\ref{lim}), while an anomalous
real term (\ref{p02}) will be added to pressure. This term is absolutely
necessary here because the third law of thermodynamics must be satisfied
(the entropy $S=0$ at zero temperature $T=0$). However, this term results in
negative entropy of tachyon Fermi gas at finite temperature (\ref{sg}).

The only possible way to satisfy the third law of thermodynamics and to
avoid imaginary quantities, is to perform integration within regular limits $%
k\in \left( 0,\infty \right) $, however, taking the tachyonic energy
spectrum in the form 
\begin{equation}
\varepsilon _k=\left\{ 
\begin{array}{cc}
\sqrt{k^2-m^2} & \qquad k\geq m \\ 
0 & \qquad k<m
\end{array}
\right.  \label{spe}
\end{equation}
and the distribution function in the form 
\begin{equation}
f_\varepsilon =\left\{ 
\begin{array}{cc}
1/\left\{ \exp \left[ \left( \varepsilon _k-\mu \right) /T\right] \pm
1\right\} & \qquad k\geq m \\ 
1 & \qquad k<m
\end{array}
\right.  \label{f2}
\end{equation}
rather than 
\begin{equation}
f_\varepsilon =\left\{ 
\begin{array}{cc}
1/\left\{ \exp \left[ \left( \varepsilon _k-\mu \right) /T\right] \pm
1\right\} & \qquad k\geq m \\ 
0 & \qquad k<m
\end{array}
\right.  \label{f22}
\end{equation}
Thus, substituting (\ref{f3}) in (\ref{n}), we have 
\begin{equation}
n=\frac \gamma {2\pi ^2}\int\limits_m^\infty f_\varepsilon \,k^2dk+\frac
\gamma {6\pi ^2}m^3  \label{nn}
\end{equation}

The distribution function (\ref{f2}) of a Fermi gas at zero temperature is
reduced to 
\begin{equation}
f_\varepsilon =\left\{ 
\begin{array}{cc}
\Theta \left( \varepsilon _F-\varepsilon _k\right) & \qquad k\geq m \\ 
1 & \qquad k<m
\end{array}
\right.  \label{f3}
\end{equation}
Substituting (\ref{spe}) and (\ref{f3}) in (\ref{e}), (\ref{pm}) and (\ref
{p0}), we find obtain the same energy density (\ref{e2}) and pressure (\ref
{p2}), while the anomalous term (\ref{p0}) vanishes at all.

According to (\ref{nn}), the particle number density is determined by
formula (\ref{n1}) rather than (\ref{n2}). As a result the entropy (\ref{s})
vanishes because $E+P-\varepsilon _Fn=0$, and, contrary to the previous
statement \cite{M84}, the third law of thermodynamics is not violated.

Formula (\ref{n1}) implies that the Fermi momentum should be defined as 
\begin{equation}
k_F=\left( \frac{\gamma n}{6\pi ^2}\right) ^{1/3}\geq m  \label{mom}
\end{equation}
rather than 
\begin{equation}
k_F=\left( \frac{\gamma n}{6\pi ^2}+m\right) ^{1/3}\geq m  \label{mom1}
\end{equation}
and that the lowest possible particle number density is 
\begin{equation}
n_{\min }=\frac \gamma {6\pi ^2}m^3  \label{min}
\end{equation}
rather than $n_{\min }=0$. However, these changes concern the very link
between the Fermi momentum $k_F$ and the particle number density $n$, while
it is not reflected in the energy density and pressure, which depend only on 
$k_F$ without regard of how $k_F$ is defined. Indeed, the sound speed is
determined by the same formula (\ref{c}), and the causality (\ref{ca}) is
satisfied under the same condition $k_F\geq \sqrt{3/2}m$ (\ref{k32}). The
ratio $P/E$ can exceed the unit, and all peculiar behavior of cold tachyon
Fermi gas \cite{TV2011c} remains valid.

The only consequence of formula (\ref{nn}) may concern the low-temperature
expansion of the Fermi level \cite{T2011h}. Let us define dimensionless
variables 
\begin{equation}
x=\frac{\varepsilon _k}T\qquad \lambda =\frac \mu T\qquad \beta =\frac mT
\label{xxx}
\end{equation}
and write formula (\ref{nn}) for a Fermi-Dirac distribution function at
finite temperature: 
\begin{equation}
n=\frac{\gamma T^3}{2\pi ^2}\int\limits_0^\infty \frac{\sqrt{x^2+\beta ^2}xdx%
}{\exp \left( x-\lambda \right) +1}+\frac \gamma {6\pi ^2}m^3  \label{nn1}
\end{equation}
An arbitrary integral 
\begin{equation}
J\left( \lambda \right) =\int\limits_0^\infty \frac{g\left( x\right) dx}{%
\exp \left( x-\lambda \right) +1}  \label{j}
\end{equation}
is expanded at low temperature ($\lambda \gg 1$) in the following series 
\cite{T2011h,Ziman} 
\begin{equation}
J\left( \lambda \right) =G\left( \lambda \right) -G\left( 0\right)
+g^{\prime }\left( \lambda \right) \frac{\pi ^2}6+g^{\prime \prime \prime
}\left( \lambda \right) \frac{7\pi ^4}{360}+...  \label{i55}
\end{equation}
So we immediately calculate 
\begin{equation}
\int\limits_0^\infty \frac{\sqrt{x^2+\beta ^2}xdx}{\exp \left( x-\lambda
\right) +1}=\frac 13\sqrt{\left( \lambda ^2+\beta ^2\right) ^3}-\frac 13{%
\beta ^3}+\frac{\pi ^2}6\frac{2\lambda ^2+\beta ^2}{\sqrt{\lambda ^2+\beta ^2%
}}  \label{j2}
\end{equation}
and, substituting (\ref{j2}) in (\ref{nn1}), we obtain 
\begin{equation}
n=\frac \gamma {6\pi ^2}q^3+\frac \gamma {12}\frac{2q^2-m^2}qT^2  \label{nn2}
\end{equation}
where 
\begin{equation}
q=\sqrt{\mu ^2+m^2} \qquad \qquad \mu\underset{T\rightarrow 0}{%
\longrightarrow }\sqrt{\varepsilon _F^2+m^2}=k_F  \label{q}
\end{equation}
Taking in to account that fact that the number of particles is conserved and
that the particle number density at zero temperature is defined by formula (%
\ref{n1}), we find 
\begin{equation}
q=k_F\left( 1-\frac{\pi ^2}6\frac{k_F^2+\varepsilon _F^2}{k_F^4}T^2\right)
\label{qz}
\end{equation}
\begin{equation}
\mu =\varepsilon _F\left( 1-\frac{\pi ^2}6\frac{k_F^2+\varepsilon _F^2}{%
k_F^2\varepsilon _F^2}T^2\right)  \label{mz}
\end{equation}
It should be noted that the same expressions (\ref{qz})-(\ref{mz}) are
derived if we redefine the particle number density by formula (\ref{n2})
because the terms with $m^3$ are mutually annihilated in expansion (\ref{j2}%
)-(\ref{nn2}) . Formula (\ref{n2}) was used in our analysis of
low-temperature expansion for the Fermi level of tachyon \cite{T2011h}.
Thus, all previous results are valid, and the entropy and specific heat of
tachyon Fermi gas are determined by formula \cite{T2011h} 
\begin{equation}
C_V=S=\frac{\gamma V}6\varepsilon _Fk_FT  \label{s6}
\end{equation}

\begin{figure}[tbp]
\caption{The particle number density $n$ of ideal gases of tachyonic thermal
excitations vs temperature variable $\beta = m/T$. Solid line: calculation
according to formula (\ref{nex}, dashed line -- according to Ref.~\cite
{TV2011d}. }
\label{ano2}{\includegraphics[scale=0.8]{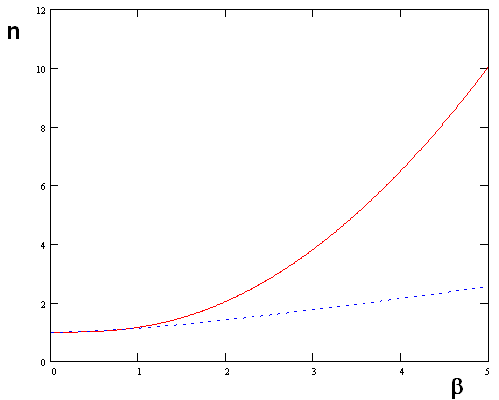}}
\end{figure}

The energy density, pressure, entropy and specific heat of tachyonic
excitations \cite{TV2011d} do remain right defined. The only correction
concerns the particle number density, which is now 
\begin{equation}
n=\frac{\gamma T^3}{2\pi ^2}\int\limits_0^\infty \frac{\sqrt{x^2+\beta ^2}%
\,xdx}{\exp x+1}+n_0=\frac{\gamma T^3}{2\pi ^2}\left( \int\limits_0^\infty 
\frac{\sqrt{x^2+\beta ^2}\,xdx}{\exp x+1}+\frac{\beta ^3}3\right)
\label{nex}
\end{equation}
where 
\begin{equation}
n_0=\frac \gamma {6\pi ^2}m^3  \label{n0}
\end{equation}
Its graph is shown in Fig.~\ref{ano2}. However, there is no qualitative
difference from the previous result (Fig.~1 in Ref. \cite{TV2011d}).

As for the hot tachyon gas \cite{T2011i}, its particle number density is now
determined so 
\begin{equation}
n=\frac{\gamma T^3}{2\pi ^2}\exp \left( -\frac \mu T\right)
\int\limits_0^\infty \sqrt{x^2+\beta ^2}\,\exp \left( -x\right) xdx+n_0
\label{nex2}
\end{equation}
and the pressure and energy density are determined so 
\begin{equation}
P=\left( n-n_0\right) T  \label{pex}
\end{equation}
\begin{equation}
E=\left( n-n_0\right) TJ\left( T\right)  \label{eex}
\end{equation}
This problem needs special analysis.

We have introduced the concept of anomalous pressure (\ref{p0}). This
pressure vanishes in a tachyon gas. However, the anomalous term (\ref{p0})
will be finite if the single-particle energy spectrum does not satisfy
condition 
\begin{equation}
\varepsilon _k\underset{k\rightarrow \infty }{\longrightarrow }\infty
\label{co1}
\end{equation}
An example of such energy spectrum is \cite{TV2011f} 
\begin{equation}
\varepsilon _k\sim \frac 1{k^\alpha}\qquad \alpha>0  \label{co2}
\end{equation}
The concept of anomalous pressure (\ref{p0}) should be considered in detail
in application to the exotic matter where it may play very significant role.

The author is grateful to Erwin Schmidt for discussions.

\newpage

\end{document}